\documentclass{ifacconf}
\usepackage{natbib} 

\usepackage{graphicx}          
\usepackage{amssymb,amsmath}
\usepackage{color}
\newcommand{\MeijerG}[7]{G \begin{smallmatrix} #1 & #2 \\ #3 &
    #4 \end{smallmatrix} \left( \begin{smallmatrix} #5 \\
      #6 \end{smallmatrix} \middle\vert #7 \right) } 

\begin{document}
\begin{frontmatter}
\title{Modeling the interference generated from car base stations towards indoor femto-cells} 
%
\thanks[footnoteinfo]{This work was supported in part by the Academy-of-Finland-funded project SMACIW under Grant no. 265040 and part of this work has been performed in the framework of the FP7 project ICT 317669 METIS, which is partly funded by the European Union. The authors would like to acknowledge the contributions of their colleagues in METIS, although the views expressed are those of the authors and do not necessarily represent the project.}


%
\author{Byungjin Cho,} 
\author{Konstantinos Koufos,} 
\author{Kalle Ruttik,}
\author{and Riku J\"antti}

\address{Aalto University, School of Electrical Engineering, Espoo, Finland 
Byungjin.cho@aalto.fi, Konstantinos.koufos@aalto.fi, Kalle.ruttik@aalto.fi, Riku.jantti@aalto.fi}

%


\begin{keyword}                           
Interference modeling, poisson point process, shared spectrum access, vehicular communication
\end{keyword}                             
\begin{abstract}                          
In future wireless networks, a significant number of users will be vehicular. One promising solution to improve the capacity for these vehicular users is to employ moving relays or car base stations. The system forms cell inside the vehicle and then uses rooftop antenna for backhauling to overcome the vehicular penetration loss. In this paper, we develop a model for aggregate interference distribution generated from moving/parked cars to indoor users in order to study whether indoor femto-cells can coexist on the same spectrum with vehicular communications. Since spectrum authorization for vehicular communications is open at moment, we consider two spectrum sharing scenarios (i) communication from mounted antennas on the roof of the vehicles to the infrastructure network utilizes same spectrum with indoor femto-cells  (ii) in-vehicle  communication utilizes same spectrum with indoor femto-cells while vehicular to infrastructure (V2I) communication is allocated at different spectrum. Based on our findings we suggest that V2I and indoor femto-cells should be allocated at different spectrum. The reason being that mounted roof-top antennas facing the indoor cells generate unacceptable interference levels. On the other hand, in-vehicle communication and indoor cells can share the spectrum thanks to the vehicle body isolation and the lower transmit power levels that can be used inside the vehicle. 
\end{abstract}
\end{frontmatter}

\section{Introduction}
The desire for entertainment and infotainment services while on board
and the need for better traffic safety is expected to increase the
traffic generated by vehicles in the near future \cite{Schneiderman,80211p}. 
For the time being infotainment services utilize the existing
macro-cellular 3GPP  infrastructure. However, in dense urban cities
and/or during traffic jams, the capacity demand can be high and the
existing infrastructure might not be able to support a fast access to
the public cloud for all users. Also, nowadays, passengers are
connected to the macro-cellular network directly, through the metallic
vehicle body that can introduce losses up to about $10\sim20$ dB \cite[p.57]{Sheikh}. In case metal coated windows are used, the losses can be even higher. Providing an adequate quality of experience for
the users on-move requires the development of new technical
enablers. Possible enablers can include ultra-dense network
deployments and mounted antennas on the roof of the vehicles acting as
moving relays \cite{Sui2013}.
   
Nowadays, cellular deployments in urban environments are characterized
by inter-site distances as low as $200$ m~\cite{Harri2012}. Further network densification
along urban streets (hereafter street micro cells) can be used to
increase the amount of load that can be served. At the same
time, passengers inside a vehicle can connect to the micro base
station through a gateway with an antenna mounted on the roof of the
vehicle overcoming high penetration losses.  

The authorization scheme for vehicular communication
is not yet standardized. One option could be to use same spectrum with
macro-cellular network, i.e. $800$ MHz to $2.6$ GHz in urban areas and
either partition the spectrum between macro and micro layer or use the
full spectrum under a shared spectrum access regime. Same approach has
been proposed for overlaying indoor femto-cells to the macro-cellular
network~\cite{Kaufman}. In this paper, we study spectrum
coexistence issues between the micro and the femto layer. We consider two different possibilities: (i) femto-cells utilize same spectrum with street micro-cells i.e. antenna on the roof of the vehicle generates interference indoors (ii) femto-cells utilize same spectrum with in-car base station while street micro-cells are allocated at different spectrum. In order to assess the impact of vehicular transmissions to indoor users, 
we develop a model for aggregate interference and SIR distribution. 

Modeling of aggregate interference level has received a lot of
attention in the literature~\cite{Haenggi_large,Cardieri,Cho1,Cho2,Ruttik}. Recently, elements from stochastic
geometry have been incorporated into interference related studies due
to their analytical tractability: When the locations of transmitters
are modeled using a Poisson point process (PPP), the Laplace transform
(LT) of the interference distribution can be expressed in
an integral form and for pathloss exponent equal to four, the integral
of the inverse LT can be solved in a closed-form~\cite{Andrews}. Usually, a
simple power law model is used 
to describe the distance-based attenuation from each individual
transmitter and the impact of fast/slow fading can also be
incorporated into the model. While power law model is sufficient to
describe distance-based propagation path loss in outdoor macro-cells,
there are more accurate models to describe attenuation along street
micro-cells. According to these models the power law changes beyond a
certain distance breakpoint~ \cite{Andersen}. This makes the analytical
treatment of interference distribution more involved. 

In this paper we compute the LT and the moments of
aggregate interference generated from parked/moving cars at the worst
case located indoor user facing a crossroad. We use the method of
moments to approximate the distribution of aggregate interference. We
show that the inverse Gamma distribution approximates closely the
interference distribution in the upper tail. Since the outage
probability at the indoor user is determined from the lower tail of
the SIR distribution, modeling the interference by inverse Gamma
distribution turns out to be a reasonable choice.  

With a low-complex and accurate interference model at hand we turn our
attention to the SIR distribution modeling. We assume Nakagami-m
channel model for radio propagation indoors and express the outage
probability at the femto-cell in terms of the Meijer G function which in our cases of interests reduces to generalized hypergeometric function widely available in most of today's software packages for numerical computations. By using it, we study how the density of cars, the uplink transmit power
level and the car isolation impact the outage probability at 
the femto-cell. We show that mounted roof-top antennas generate unacceptably  high interference level at the femto-cell while in-car and femto-cell communication can coexist in same spectrum. 
Also, we utilize the proposed
model to illustrate that the most significant part of the interference
comes from the street facing the femto-cell. Based on this
observation, we argue that for mounted roof-top antennas frequency planning between street
micro-cells and indoor femto-cells could be an option for performance
enhancement.  
\section{System model}
\begin{figure}
\begin{center}
\includegraphics[height=5cm]{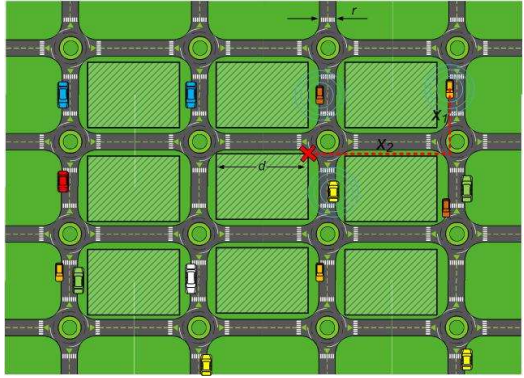}    
\caption{System illustration.}  
\label{fig:fig0}
\end{center}                                 
\end{figure}

We consider a dense urban city with street layout following the
Manhattan grid. The building 
block length is denoted by $d$ and the street width is denoted by 
$r$. Let us define the distance  
between neighboring streets as $D\!=\!d\!+\!r$, see Fig.~\ref{fig:fig0}. Along  
a street, the vehicles are distributed according to a PPP model. At  
particular snapshot, each vehicle is active in the uplink with a
certain probability. As a result, the distribution of uplink
transmissions along a street  
follows a PPP too. 
We denote the PPP on the $k$-th vertical street by
$\Phi_k$ with density (i.e. number of uplink transmissions per meter) 
$\lambda_k$  and on the ${k'}$ th horizontal street by $\Phi_{k'}$ with
density $\lambda_{k'}$. 

The uplink transmit power level is denoted by  $P_t$ and 
the fading channel $h$ for vehicular transmissions is assumed to
be Rayleigh with mean equal  
to unity, $E\{h \} \!=\! 1$. The distance-based pathloss in Manhattan grid, 
$l(x_1,x_2)$, is a function of the NLOS distance, $x_1$, as well as of
the LOS distance $x_2$, see also Fig.~\ref{fig:fig0}. For a vehicle located at
coordinates $(x_1,x_2)$, the distance-based pathloss to the origin 
can be read as $l(x_1,x_2)=C x_1^{-2} x_2^{-\alpha}$ where the
attenuation exponent $\alpha$ along the 
LOS is taken equal to $2$ and typical values for the NLOS are
$4\sim 6$. One may notice that for the $k$-th vertical street $x_{2,k}=kD,k>0$
while for the vertical street facing the considered femto-cell, $k=0$,
we have $x_{2,k}=1$. Also, the parameter $C$ accounts for antenna
gain, car isolation (in the case of in-car base station) and
wall losses with typical wall attenuation in the range of $15\sim20$
dB. 

We want to identify whether uplink vehicular communication and indoor
femto-cells can coexist in the same spectrum. In order to do that we
evaluate the outage probability for given SIR target
$\gamma$ at the worst-case located femto-cell facing a 
cross-road, see Fig.~1. In order to evaluate the SIR distribution we
need a model for the aggregate interference distribution and a model
for the wanted signal power level at the femto-cell. Assuming
Nakagami-m fading for indoor propagation, the wanted signal power is
Gamma distributed with shape $m$ and scale ${\overline{P}}_{rx}/m$
where ${\overline{P}}_{rx}$ denotes the mean received signal
power. Next, we propose a model for the interference distribution and
evaluate its moments. 
\section{Modeling interference distribution}
The interference level 
can be expressed as the sum of interference levels from all active
car base stations on all vertical and horizontal streets. 
\begin{eqnarray}  
I_{\Phi} &{=}& \sum_{k}I_{\Phi_k} + \sum_{k'}I_{\Phi_{k'}}\nonumber\\ 
 &{=}& \sum_{k}\!\!\sum_{x_{i}\in \Phi_k} \!\!\!\!P_{t} h_i l(x_{1,i},x_{2,k}) \!+\!
 \sum_{k'}\!\!\sum_{x_{i'}\in \Phi_{k'}} \!\!\!\!P_{t} h_{i'}
 l(x_{1,i'},x_{2,{k'}})\nonumber 
\end{eqnarray} 
where $x_{1,i}$ and $x_{2,k}$ are
vertical NLOS distance and horizontal LOS distance, respectively,
between $i$-th car on $k$-th vertical street to the crossing toward
the victim building, and $x_{1,i'}$ and $x_{2,{k'}}$ are horizontal NLOS
distance and vertical LOS distance, respectively,  between ${i'}$-th
car on ${k'}$-th horizontal street to the crossing toward the victim
building. 

Based on the assumption that
vertical/horizontal streets in Manhattan 
two-dimensional grid are symmetric and a constant car 
density is used on all the streets, we have same density for the
processes $\Phi_k, \Phi_{k'} \forall k,k'$, or, 
$\lambda_k\!=\!\lambda_{k'}\!=\!\lambda\, \forall k,k'$. Also, the
independent property of PPP  
allows us to focus only on one type of street (hereafter, vertical
streets) and incorporate the impact of horizontal streets using a
scaling factor equal to two. Next, we characterize the distribution of
aggregate interference from uplink transmissions located on the $k$-th vertical
street, $I_{\Phi_{k}}$, in terms of its LT. 
\begin{lem}
For Rayleigh fading channels, i.e. $h_i \sim \exp(1)$, the LT of the
aggregate interference from car base stations 
distributed as one-dimensional PPP on the $k$-th
vertical street in a Manhattan grid is 
\begin{eqnarray} \label{lem2}
\mathcal{L}_{I_{\Phi_k(z)}}(s) & {=}& \mbox{exp}\left[ -
  \pi\lambda \sqrt{\frac{sz}{\beta_k}}  \right]
\end{eqnarray}
where  $z = P_{t}C$ and $\beta_k\!=\!x_{2,k}^{\alpha}$. The proof is given in Appendix. 
\end{lem}

\begin{prop}
The total interference from all
vertical streets in a Manhattan grid is  
\begin{eqnarray} 
\label{eq:prop1}
\mathcal{L}_{I_{\Phi(z)}}(s)  &{=}& \mbox{exp}\left[ -\pi\lambda
  \left\lbrace 1 + { \zeta\left(\frac{\alpha}{2}\right)}{}
    D^{-\frac{\alpha}{2}}\right\rbrace\sqrt{sz}\right]
\end{eqnarray} 
where $\zeta(\cdot)$ denotes the Riemann zeta function. The proof is given in Appendix. 
\end{prop}

Note that the integral range (refer to the Proof) used in Eq.(\ref{lem2}) is from $0$ to
infinity, i.e. singular pathloss model. This range causes inaccurate
interference modeling in 
the near-field, since the interference level can become infinite when
$x_{1}$ and $x_{2,k}$ become zero. In a more practical case, non-singular
path loss model is taken into account by limiting $x_1 \geq 1$, and
apply the integral $\int_1^\infty (\cdot)dx_1$.   
The results given in Eq.~(\ref{lem2}) and Eq.~\eqref{eq:prop1} can be
regenerated through the following Lemma and Proposition. 
\begin{lem}
For the non-singular path loss model, the interference from the $k$-th 
vertical street in a Manhattan street is 
\begin{eqnarray}  
\mathcal{L}_{I_{\Phi_k(z)}}(s) &{=}& \mbox{exp}\left[ - 2\lambda
  \sqrt{\frac{sz}{\beta_k}}
  \arctan\left(\sqrt{\frac{sz}{\beta_k}}\right)\right]. \nonumber
\end{eqnarray}
The proof is given in Appendix.
\end{lem}
\begin{prop}
For the non-singular path loss model, the total interference from all
vertical streets in a Manhattan grid is  
\begin{eqnarray}  
\label{eq:prop2}
\mathcal{L}_{I_{\Phi(z)}}(s) &{=}& \mbox{exp}\left[ - 2\lambda
  \sum_{k=0}^{\infty}\sqrt{\frac{sz}{\beta_k}}
  \arctan\left(\sqrt{\frac{sz}{\beta_k}}\right)\right]. 
\end{eqnarray} 
\end{prop}
The integral for the inverse LT in Eq.~\eqref{eq:prop2} does not exist
in closed-form. 
As a means to provide simple and useful expressions, the use of
approximations for the aggregate interference is motivated. Since the LT of the interference with the singular
pathloss model in Eq.~\eqref{eq:prop1} resembles the Characteristic function
of the Levy distribution, we approximate the interference by a fitted
inverse Gamma distribution.

The first two moments of the interference can be found based on the LT
$\mathbf{E}(I_{\Phi}^i) = (-1)^i \frac{d^i }{ds^i}
\mathcal{L}_{I_{\Phi}}(s)\vert_{s=0}$ and expressed as following property of the
\begin{eqnarray}
\mathbf{E}\{I_{\Phi}\} \!&{=}&\!  2\lambda \!\sum_{k=0}^{\infty} {\frac{z}{\beta_k}}
{=}  2\lambda z \left(1+\frac{\zeta(\alpha)}{ D^{\alpha} }\right) \nonumber\\ 
\mathbf{E}\{I_{\Phi}^2\} \!&{=}&\!  \frac{4\lambda}{3} \!\sum_{k=0}^{\infty}
\!{\frac{z^2}{\beta_k^2}}\! + \!\mathbf{E}\{I_{\Phi}\}^2  {=}   \frac{4\lambda z^2}{3}  \!\!\left(\!1\!+\!\frac{\zeta(2\alpha)}{
    D^{2\alpha} }\!\right) \!+\! \mathbf{E}\{I_{\Phi}\}^2.\nonumber
\end{eqnarray}
The fitted inverse Gamma distribution has 
scale $a = \frac{\mathbf{E}\{I_{\Phi}\}^2}{\mathbf{E}\{I_{\Phi}^2\} - \mathbf{E}\{I_{\Phi}\}^2} + 2$ and shape
$b = {\mathbf{E}\{I_{\Phi}\}}{(a - 1)}$. 
The
inverse of the interference distribution is a Gamma distributed random 
variable with scale $a$ and shape $1/b$, $I_{\Phi}^{-1} \sim {\text{Gamma}}(a, 1/b)$. 

\section{Modeling SIR distribution}
For the performance evaluation of wireless systems, assessing the
quantity of the signal to interference plus noise ratio (SINR) is
required. Based on our system model assumptions, the performance at
the worst case femto-cell is limited by 
the interference level and the impact of noise power can be
ignored. The SIR level is 
\begin{eqnarray}
{SIR} = \frac{P_{rx}}{I_{\Phi}} \nonumber
\end{eqnarray}
where $P_{rx}$ is the wanted received signal level. 

For ensuring satisfactory services to users of the femto-cell network,
the outage probability (i.e. probability that the SIR value is below a
target SIR)  must be maintained under specific value. The outage
probability can be computed using the following Lemma. 
\begin{lem}
The outage probability when the source destination fading is
Nakagami-m distributed, with mean signal level ${\overline{P}}_{rx}$
is 
\begin{eqnarray}
\label{eq:lem4}
\mathbf{P}({SIR}\leq \gamma) &=& 1- \sum_{i=0}^{m-1}
\frac{(-\xi)^i}{i!}  \frac{d^i}{d\xi^i} \mathcal{L}_{I_{\Phi(z)}}(\xi)
\end{eqnarray}
where $\xi = \frac{m\gamma}{\overline{P}_{rx}}$.  
The proof is given in the Appendix. 
\end{lem}


Eq.~\eqref{eq:lem4} indicates that the LT of the aggregate
interference generated by the car base stations  
plays a central role in quantifying the outage probability in the 
femto-cell network. While the outage probability is completely
characterized by the LT of the interference, it is also useful to
approximate the SIR distribution by some known function so as to
assess its mean and higher moments in a low-complex manner. 

Note that the useful signal level at the femto-user in the Nakagami-m
fading channel is Gamma distributed with scale $m$ and shape $\theta =
\frac{\overline{P}_{rx}}{m}$, $P_{rx} \sim $ Gamma($m, \theta$). Hence, with
$I_{\Phi}^{-1} \sim {\text{Gamma}}(a, 1/b)$ presented in Section 3,
the SIR, $P_{rx}\cdot I_{\Phi}^{-1}$, can be expressed as product of two
independent Gamma random variables. By normalizing respective shape parameters,
the normalized SIR $\gamma_n = \frac{b}{\theta}\cdot\frac{
  P_{rx}}{I_{\Phi}}$ becomes the product of two Gamma random variables with
shape equal to unity, $g_1 = P_{rx} \theta^{-1} \sim$  Gamma($m, 1$) and
$g_2 = I_{\Phi}^{-1}{b} \sim$  Gamma($a, 1$) and $\gamma_n = g_1\cdot
g_2$. In that case the PDF of the normalized SIR distribution can be
expressed in terms of the Meijer G function\cite{Springer}
\begin{eqnarray}  
\mathbf{P}\left(\gamma_n\right) &=&
\frac{\MeijerG{2}{0}{0}{2}{-}{m,a}{\gamma_n}}{\Gamma(m)\Gamma(a)} 
= \frac{2{\gamma_n}^{\frac{a+m}{2} }\cdot K_{m-a}(2\sqrt{\gamma_n} )}{\Gamma(m)\Gamma(a)}   \nonumber
\end{eqnarray} 
where $\Gamma(\cdot)$ is the gamma function, $G(\cdot)$ is Meijer G-function and 
$K_{m-a}(\cdot)$ is the modified Bessel function of the second kind with order ($m-a$).

The CDF of the normalized SIR distribution is derived
using integral properties \cite{Springer} of the Meijer G-function, 
\begin{eqnarray}  \label{eq:cdf}
\mathbf{P}\left(\frac{b}{\theta}\cdot\frac{ S}{I_{\Phi}} \leq \gamma_n\right) &=&
\frac{\MeijerG{2}{1}{1}{3}{1}{m,a, 0}{\gamma_n}}{\Gamma(m)\Gamma(a)}  
\end{eqnarray} 
where
$\MeijerG{2}{1}{1}{3}{1}{m,a, 0}{\gamma_n} =\frac{\Gamma(a-m)
  \gamma_n^m}{m} {}_1F_2(m;m+1-a,m+1;\gamma_n) + \frac{\Gamma(m-a)
  \gamma_n^{a}}{a} {}_1F_2(a;a+1-m,a+1;\gamma_n) $ and ${}_pF_q(\cdot)$ 
  is the generalized hypergeometric function for integer $p$ and $q$.

Eq. (\ref{eq:cdf}) can be used as the CDF of the SIR distribution by
properly shifting the axis
$\mathbf{P}\left(\frac{b}{\theta}\cdot\frac{ S}{I_{\Phi}} \leq
  \gamma_n\right) = \mathbf{P}\left(\frac{ S}{I_{\Phi}} \leq
  \frac{\theta}{b}\gamma_n\right) = \mathbf{P}\left(\frac{ S}{I_{\Phi}} \leq
  \gamma\right)$.

\section{Numericals}
In order to verify the suitability of interference and SIR models
presented in Section 3 and Section 4 we consider cars deployed in a
city centre which is modeled as a square with side equal to $2$ km. Street
width is taken equal to $r\!=\!20$ m and building block is a
square with side $d\!=\!50$ m.  For NLOS propagation the pathloss 
exponent is taken equal to $\alpha\!=\!4$. {\color{black}The transmit power level in the
uplink is $P_t = 100$ mW, the wall attenuation is $15$ dB, the attenuation
constant is $10^{-3}$. For mounted roof-top antenna, car isolation is equal to $\eta=0$ dB and for in-car base station  $\eta= \{10, 20, 30\}$ dB are considered. For
indoor propagation we present results for Rayleigh fading, $m=1$.} The mean
wanted signal level at the femto-cell is taken
equal to $\overline{P}_{rx} = -40$ dBm. In order to evaluate the outage probability at the
femto-cell we assume a SIR target equal to $\gamma = 15$ dB. 

\begin{figure}
\begin{center}
\includegraphics[height=6cm]{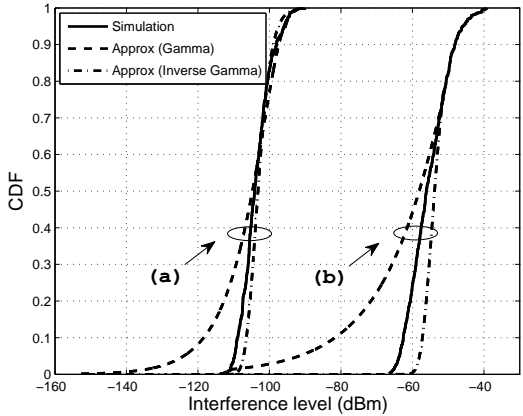}    
\caption{Approximations of interference level distribution using Gamma
  and Inverse-Gamma distribution with non-singular path-loss model,
  when (a) the interference from the closest street $k=0$ is not
  considered and (b) the interference from all vertical streets is
  considered.}  
\label{fig:fig1}                                 
\end{center}                                 
\end{figure}

Firstly, we check the accuracy of fitted inverse Gamma distribution for
approximating the aggregate interference. For
the parameter settings presented above the scale and the shape of the
fitted distribution are $a=2.3$ and $b=8.2\cdot 10^{-9}$
respectively. Recall that the 
inverse Gamma approximation was selected on the basis of the LT of the
interference for singular pathloss model in Eq.~\eqref{eq:prop1}
which resembles the characteristic function of the Levy
distribution. For comparison purposes we 
illustrate the accuracy for a fitted Gamma distribution too, see
Fig.~\ref{fig:fig1}. Both 
distributions are accurate in the upper tail which would determine the
accuracy of the approximation in the lower tail of the SIR
distribution. However, the inverse Gamma distribution achieves better
approximation over the full distribution body. In Fig.~\ref{fig:fig1}, it is
interesting to note that the mean interference level decreases by
approximately $50$ dB if the impact of the street facing the femto-cell
of interest, $k=0$, is not considered. 
\begin{figure}
\begin{center}
\includegraphics[height=6cm]{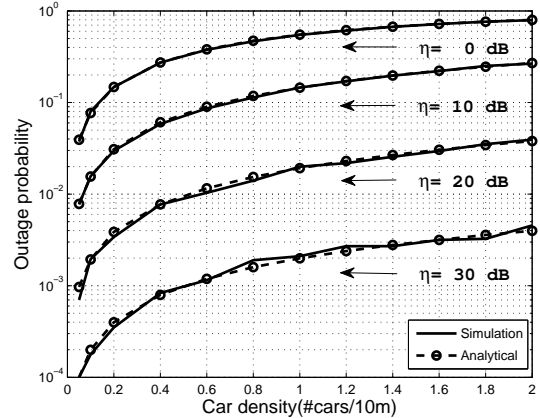}
\caption{Comparison of outage probability for simulation result and
  analytical result in Eq.~\eqref{eq:lem4}  with respect to car base
  station density $\lambda$ and car isolation $\eta$ when Nakagami fading parameter $m = 1$, mean wanted
  signal level is $\overline{P}_{rx} = -40$ dBm  and SIR target is $\gamma = 15$ dB.}
\label{fig:fig2} 
\end{center}                                
\end{figure}
\begin{figure}
\begin{center}
\includegraphics[height=6cm]{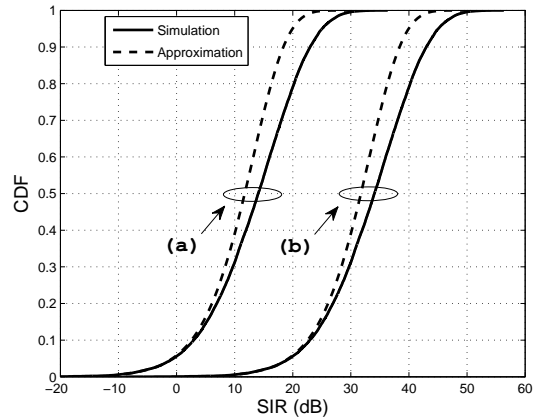}    
\caption{SIR distribution when car base
  station density is $\lambda=0.1$ Cars/m, the mean wanted signal level is $\overline{P}_{rx} = -40$ dBm, Car isolations are (a) $\eta=0$ dB and (b) $\eta=20$ dB}  
\label{fig:fig3}
\end{center}                                 
\end{figure}

Secondly, we utilize Eq.~\eqref{eq:lem4} for approximating the outage
probability at the femto-cell  as a function of the density $\lambda$ of uplink
transmissions, see Fig.~\ref{fig:fig2}. 
With car isolation equal to $20$ dB or higher, the outage probability decreases at acceptable values even for a high density of vehicles. 


The outage probability has been approximated by evaluating the LT of
the interference distribution and also its derivatives at certain
point, see Eq.~\eqref{eq:lem4}. The amount of computations might be
high particularly when the shape $m$ of the Nakagami distribution
is high. The CDF of the SIR distribution has been approximated in
Section 4 using the Meijer G function. The approximating CDF can be
useful for estimating not only the outage probability but also the
moments of the SIR distribution. In Fig.~\ref{fig:fig3} one can
observe that the Meijer G function is accurate in the lower
tail. However, the mean SIR is underestimated by $2.5$ dB. 


\section{Conclusion}
In dense urban cities vehicular
communication may use same spectrum with indoor
femto-cells. As a result, better quality of experience for the 
end-users while on board would most probably come at the cost of
higher interference generated indoors. We developed a model that is useful
for assessing the outage probability at the
femto-cell due to the interference generated from car base
stations. With mounted antennas on the top of the vehicles the outage
probability becomes prohibitively high given that the  
density of vehicular transmissions is high too. 
On the other hand, with in-car base stations, the isolation due to the car shell and the possibility to use lower power levels inside the car make it possible to maintain a low outage probability at the femto-cell even for a high car density. However, in that case, additional spectrum resources are needed as only in-car communication uses same spectrum with indoor femto-cells while street micro-cells are allocated at different spectrum. 

\appendix
\section*{APPENDIX}   
\section{Proof of Lemma 1}   
The distribution of the random variable $I_{\Phi_k}$ is characterized
in terms of its LT which is given by  
\begin{eqnarray}
\mathcal{L}_{I_{\Phi_k(z)}}(s) &=& \mathbb{E}\left[{e}^{-s\cdot
    I_{\Phi_k}}\right] =  
\mathbb{E}\left[{e}^{-s\cdot \sum_{x_i\in\Phi_k} P_t h_{x_i}
    l(x_i,x_{2,k})}\right]\nonumber\\ 
&\overset{(p1)}{=}& \mathbb{E}_{\Phi_k}\left[\prod_{x_i\in\Phi_k}
  \mathbb{E}_h [{e}^{-s\cdot P_t h_{}
    l(x_i,x_{2,k})}]\right]\nonumber\\ 
&\overset{(p2)}{=}& {e}^{ - \lambda \int_{\mbox{${\mathbb
        R}$}} \left(1- \mathbb{E}_h \left[{e}^{-s\cdot P_t h_{}
        l(x,x_{2,k})}\right]\right)\mbox{d}x } \nonumber\\ 
&\overset{(p3)}{=}&  {e}^{ - 2\lambda \int_0^{\infty}
  \left(1-{\mathbb{E}_h \left[{e}^{-s\cdot P_t h
          l(x,x_{2,k})}\right]}_{}
  \right)\mbox{d}x}\nonumber\\ 
&\overset{(p4)}{=}& {e}^{ - 2\lambda \int_0^{\infty}
  \frac{s\cdot P_t l(x,x_{2,k})}{1 + s\cdot P_t l(x,x_{2,k})}
  \mbox{d}x}\nonumber	\\ 
& \overset{(p5)}{=}& {e}^{- 2\lambda \int_0^{\infty}
  \frac{s z}{\beta_k x^2 + s z} dx} \nonumber   
\overset{(p6)}{=} {e}^{ - \pi\lambda
  \sqrt{\frac{sz}{\beta_k}}}   \nonumber 
\end{eqnarray} 

where ($p1$) follows from the i.i.d. distribution of the fading $h$,
($p2$) follows from the generating functional \cite{Haenggi_large} of the one-dimensional
PPP, ($p3$) follows from the symmetry of the interference around the
origin, and ($p4$) follows from the fact that the fading $h$ follows an
exponential distribution with mean equal to unity. ($p5$) follows the pathloss model
$l(x,x_2)=Cx^{-2}x_2^{-\alpha}$ and ($p6$) follows
from the integration rule $\int_0^\infty \frac{s}{ux^2 + s} \mbox{d}x =
\frac{\pi}{2}\sqrt{\frac{s}{u}}$ \cite[2.172]{Zwillinger}.


\section{Proof of Proposition 2} 
Since the PPP:s along different vertical streets are independent among
each other, the LT of the aggregate interference from all vertical
streets is equal to the product of the LT:s from each vertical street 
\begin{eqnarray}
\mathcal{L}_{I_{\Phi(z)}}(s) &{=}& \prod_k\mathcal{L}_{I_{\Phi_k(z)}}(s) 
{=} {e}^{- \pi\lambda \sum_{k=0}^{\infty}
  \sqrt{\frac{sz}{\beta_k}} }  \nonumber \\ 
 &\overset{(p1)}{=}& {e}^{ -\pi\lambda  \left( 1 + {
       \zeta\left(\frac{\alpha}{2}\right)}{}
     D^{-\frac{\alpha}{2}}\right)\sqrt{sz}}
\end{eqnarray} 
where ($p1$) follows from the Riemann zeta function, $\zeta(s) =
\sum_{k=1}^{\infty} k^{-s}$ \cite[0.233]{Zwillinger}.  
\section{Proof of Lemma 3}   
For the non-singular pathloss model, the interference from $k$-th
vertical street is derived by following similar steps to Lemma
2. Only the lower integration limit is different. 
\begin{eqnarray} \label{e1}
\mathcal{L}_{I_{\Phi_k(z)}}(s) &{=}& {e}^{- 2\lambda
  \int_1^{\infty}  \frac{s\cdot P l(x, x_{2,k})}{1 + s\cdot P l(x,
    x_{2,k})} dx} \nonumber  	= 
 {e}^{ - 2\lambda \int_1^{\infty}
  \frac{s z}{\beta_k x^2 + s z} dx} \nonumber   \\ 
& \overset{}{=}& {e}^{- 2\lambda \int_1^{\infty}  \frac{1}{
    (\frac{\beta_k}{sz})x^2 + 1} dx} \nonumber   
\overset{(p1)}{=} {e}^{ - 2\lambda
  \sqrt{\frac{sz}{\beta_k}}
  \arctan\left(\sqrt{\frac{sz}{\beta_k}}\right)}
\end{eqnarray}
where ($p1$) follows from the integration rule $ \int_1^\infty
\frac{s}{ux^2 + s} \mbox{d}x = 
\sqrt{\frac{s}{u}}\cdot\mbox{arctan}\left(\sqrt{\frac{s}{u}}\right)$ \cite[2.172]{Zwillinger}. 

\section{Proof of Lemma 5} 
\begin{eqnarray}
\mathbf{P}(\mbox{SIR}\geq \gamma) &=& \mathbf{P}\left({  h \overline{P}_{rx} }/{I_{\Phi}}\geq \gamma\right) = \mathbf{P}\left( h \geq {\gamma  I_{\Phi}}/{\overline{P}_{rx}} \right) \nonumber\\
 &\overset{(p1)}{=}& \int_0^{\infty} \mathbf{P}\left( h \geq
   {\gamma t}/{\overline{P}_{rx}} \right) f_{I_{\Phi}}(t) \mbox{d}t
 \nonumber\\ 
  &\overset{(p2)}{=}& \int_0^{\infty} \sum_{i=0}^{m-1}
  \frac{(mxt)^i}{i!} e^{-mxt} f_{I_{\Phi}}(t) \mbox{d}t  \nonumber\\ 
        &\overset{(p3)}{=}& \sum_{i=0}^{m-1} \frac{(-mx)^i}{i!} 
        \frac{d^i}{d(mx)^i} \int_0^{\infty} e^{-mxt} f_{I_{\Phi}}(x)
        \mbox{d}x  \nonumber\\ 
           &=& \sum_{i=0}^{m-1} \frac{(-\xi)^i}{i!}
           \frac{d^i}{d\xi^i} {\mathcal{L}}_{I_{\Phi}}(\xi)  \nonumber 
\end{eqnarray}
where $x = \frac{\gamma}{\overline{P}_{rx}}$ and $\xi = mx$.
Equality ($p1$) follows from 
denoting $f_{I_{\Phi}}(t)$ the PDF of the interference distribution,
($p2$) is based on the fact that when the source-destination fading is
Nakagami-m distributed, then the CCDF of $h$ is $\mathbf{P}(h > x) =
\sum_{i=0}^{m-1} \frac{(mx)^i}{i!} e^{-mx}$, and ($p3$)
holds true according to the following Laplace transform property 
$\int_0^{\infty} e^{-st} t^i f_{I_{\Phi}}(t) \mbox{d}t 
=(-1)^i\frac{d^i}{ds^i}   {\mathcal{L}}\{ f_{I_{\Phi}}\}(s)$.

\end{document}